\documentclass[twocolumn,showpacs,superscriptaddress,aps,pra]{revtex4}
\usepackage{epsfig}
\usepackage{subfigure}
\usepackage{amssymb}
\usepackage{graphicx}
\usepackage{color}
\usepackage{amsfonts}
\usepackage{amscd}
\usepackage{hyperref}
\usepackage{amsmath}
\usepackage{epstopdf}
\usepackage[normalem]{ulem}

\begin{document}

\newcommand{\cmt}[1]{{\textcolor{green}{[#1]}}}
\newcommand{\qn}[1]{{\textcolor{red}{ (?)  #1 }}}
\newcommand{\chk}[1]{{\textcolor{blue}{#1}}}
\newcommand{\del}[1]{{\textcolor{blue}{ \sout{#1}}}}
\newcommand{\rvs}[1]{{\textcolor{red}{#1}}}
\newcommand{\rpl}[2]{{\sout{#1}}{\color{blue}{#2}}}        

\title{Coherent perfect absorption in a weakly coupled atom-cavity system}

\author{Wei Xiong}


\affiliation{Institute of Advanced Manufacturing Engineering, Hefei University, Anhui 230022, China}
\affiliation{Department of Applied Physics, Hong Kong Polytechnic University, Hung Hom, Hong Kong, China}
\affiliation{Interdisciplinary Center of Quantum Information and Zhejiang Province Key Laboratory of Quantum Technology and Device, Department of Physics and State Key Laboratory of Modern Optical
	Instrumentation, Zhejiang University, Hangzhou 310027, China}

\author{Jiaojiao Chen}
\affiliation{Hefei Preschool Education College, Anhui 230013, China}

\author{Baolong Fang}

\altaffiliation{fbl@hfuu.edu.cn}
\affiliation{Institute of Advanced Manufacturing Engineering, Hefei University, Anhui 230022, China}

\author{Chi-Hang Lam}

\altaffiliation{C.H.Lam@polyu.edu.hk}
\affiliation{Department of Applied Physics, Hong Kong Polytechnic University, Hung Hom, Hong Kong, China}

\author{J. Q. You}

\altaffiliation{jqyou@zju.edu.cn}
\affiliation{Interdisciplinary Center of Quantum Information and Zhejiang Province Key Laboratory of Quantum Technology and Device, Department of Physics and State Key Laboratory of Modern Optical
	Instrumentation, Zhejiang University, Hangzhou 310027, China}

\date{\today }

\begin{abstract}
We study coherent perfect absorption (CPA) theoretically based on a weakly coupled atom-cavity system with an optically pumped second-order nonlinear crystal (SOC) embedded in the cavity. Our system does not require a strong coupling, which is often needed for CPA in previous studies but is challenging to implement experimentally in some systems. The role of the SOC is to introduce a tunable  effective decay rate of the cavity, which can lead to CPA in the weak coupling regime. The proposed system exhibits bistable behaviors, with bistable patterns switchable between conventional and unconventional shapes. By varying the properties  of the SOC, the operation point of CPA can be tuned to be inside or outside the bistable regime. It can also be located at the upper or the lower stable branch or even the unstable branch of the bistable hysteresis loop. It is however robust against the parameters of the SOC for any fixed effective decay rate. Our system can potentially be applied to realize optical devices such as optical switches in the weakly coupled regime.
\end{abstract}


\maketitle
\section{Introduction}
It is well known that an optical cavity with gain can produce outgoing optical fields with a definite frequency and phase relationship, termed laser~\cite{siegmann}. Applying time reversal symmetry, a cavity illuminated by two coherent incoming waves has the gain medium replaced
by an absrobing medium. Once the coherent incoming waves are completely absorbed, coherent perfect absorption (CPA)~\cite{ydc1} occurs. Hence, CPA can be regarded as the time reversed process of lasing at the threshold~\cite{ydc1,ww}. The underlying physics of CPA is a joint action of the system disspation and the destructive interference between the transmitted and reflected fields~\cite{ydc1,hn}. Due to wide potential applications~\cite{ads,mk,mc} in optical communications and photonic devices such as transducers, modulators, optical switches and transistors, CPA has attracted considerable interest~\cite{ydc1,ww,sdg,hn,gsd,jwy,sl,sl2,ydc2,ydc3,sh,jt,xby,yml2018,ag1,ag2,ag3,tr,yz,ag4,dkz}.

Previous studies~\cite{ag1,ag2,ag3,sl2011} have shown that strong coupling between light and matter is necessary for realization of CPA and the conditions under which CPA occurs  cannot be fine-tuned. However, achieving the strong coupling regime is still a challenge for relevant quantum systems such as atom/spin-cavity systems, and a tunable system is always desirable in quantum computation and quantum information processing. Motivated by these, we put forward a proposal of a system exhibiting controllable CPA in the weak coupling regime. The proposed system consists of a two-level atom coupled to a cavity or specifically-designed cavity~\cite{hchoi} containing a second-order nonlinear crystal (SOC). The SOC gives rise to tunable effective decay rate of the cavity. This extends the CPA conditions in~\cite{ag1,ag2,ag3} to the weak coupling regime. In addition, we show that the proposed system under the new CPA conditions exhibits bistable behaviors. The properties of bistability have been widely applied in optical switching\cite{vr1,vr2}, optical memory\cite{liy,cr}, quantum phase transition\cite{lr,lintian}, ground-state cooling\cite{vmy}, laser\cite{gq}, quantum circuits\cite{tch,teo,ds,bd}. The bistable property can be switched from conventional to unconventional forms via the tuning of system parameters. Moreover, the location of the CPA point can be fine-tuned to be inside or outside the bistable region, and it can locate at an arbitrary branch of the bistable hysteresis loop. Tuning CPA appearing on different branches can be used to study nonreciprocal circulators\cite{ls}. Surely, CPA appearing on different bistable branches indicate different input powers are needed to observe CPA point. This greatly relaxes experimental conditions to observe it. We also show that the location of the CPA point is robust against the parameters of the SOC for a given effective decay rate. 

The paper is organized as follows. In Sec. II, the model is formulated theoretically and the Hamiltonian is explained. Then we apply quantum Langevin equations to derive steady-state properties of the system in Sec. III. In Sec. IV, the CPA criterion is given and impacts of the system parameters are discussed. In Sec. V, we give a numerical study on the CPA behaviors. Finally, a conclusion is given in Sec. VI.

\section{Model and Hamiltonian}

\begin{figure}
	\includegraphics[scale=0.6]{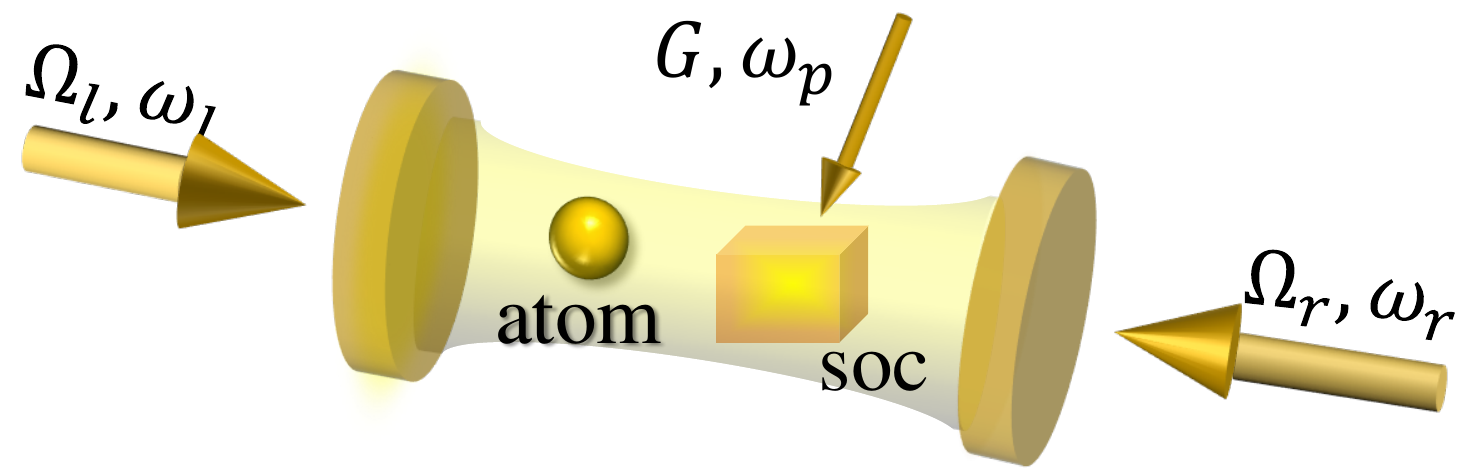}
	\caption{(Color online) Schematic diagram of the coupled atom-cavity system. The cavity containing a pumping SOC is driven by two coherent fields with Rabi frequencies $\Omega_l$ and $\Omega_r$. $G$ is the coefficient of SOC and $\omega_s~(s=l,r,p)$ is the frequency of the field.}\label{fig1}
\end{figure}

We consider a quantum system consisting of a two-level atom coupled to an optical cavity containing a SOC. The cavity is formed by two partially transmitting (reflecting) mirrors. Each mirror is exerted by an external driving field~ [see Fig.~\ref{fig1} ]. The system Hamiltonian can be written as  (setting $\hbar=1$)
\begin{align}
H_{\rm sys}=H_0+H_I+H_{\rm nl}+H_d\label{eq1}.
\end{align}
Here, $H_0=\omega_c c^\dag c+\omega_{\rm TLS}\sigma_z$ is the total free energy of the cavity and the two-level atom, where $\omega_c$ is the angular frequency of the cavity mode and $\omega_a$ is the transition frequency between the ground state $|g\rangle$ and the first excited state $|e\rangle$ of the two-level atom. Also, $c^\dag~(c)$ is the creation (annihilation) operator of the cavity field, and $\sigma_z=\frac{1}{2}[\sigma_+,\sigma_-]$ is the Pauli-$z$ operator, where $\sigma_+=|e\rangle\langle g|~(\sigma_-=|g\rangle \langle e|)$ is the raising (lowering) operator of the atom.

The term $H_I$ in Eq.~{\ref{eq1}} represents the  interaction Hamiltonian between the cavity and the two-level atom.  Under the rotating wave approximation (i.e., neglecting the fast-oscillating terms $c^\dagger\sigma_+$ and $c\sigma_-$), $H_I$ reads
\begin{align}
H_I=g (c^\dag\sigma_-+c\sigma_+),\label{eq2}
\end{align}
where $g=d\sqrt{\omega_c/(2\pi\epsilon_0 V_0)}$ is the coupling strength with $d$ being the dipole momentum, $\epsilon_0$ the vacuum permittivity, and $V_0$ the cavity mode volume.

The Hamiltonian $H_{\rm nl}$ in Eq.~(\ref{eq1}) denotes the interaction between the SOC and the cavity.  By pumping an external field onto the SOC, the nonlinear interaction Hamiltonian $H_{\rm nl}$ can be written as
\begin{align}
H_{\rm nl}=i(G c^{\dag2} e^{-i\omega_p t}-G^* c^2 e^{i\omega_p t}),\label{eq3}
\end{align}
where the parameter $G$ is the effective nonlinear coefficient, proportional to both the original nonlinear coefficient of the SOC and the amplitude of the pumping field with frequency $\omega_p$ and $G^*$ is the complex conjugate of $G$. The Hamiltonian in Eq.~(\ref{eq3}) can be used to study the phenomena of squeezing effects~\cite{zu0,zu1,zu2}  and coupling amplification~\cite{wx,wq,xylu}.

The last term $H_d$ in Eq.~(\ref{eq1}) describes the mirrors of the cavity driven by two external fields with frequencies $\omega_l$ and $\omega_r$, respectively (see Fig.~\ref{fig1}). The Hamiltonian $H_d$ takes the form of
\begin{align}
H_d=i(\Omega_l e^{-i\omega_l t}+\Omega_r  e^{-i\omega_r t})c^\dag+{\rm H.c.}, \label{eq4}
\end{align}
where $\Omega_{l}=\sqrt{\kappa_{l}} c^{(\rm in)}_{l}$ and $\Omega_{r}=\sqrt{\kappa_{r}} c^{(\rm in)}_{r}$ are the Rabi frequencies of the left and right driving fields, respectively. Here, $\kappa_{l(r)}=T_{l(r)}/\tau$  is the decay rate of the left (right) mirror of the cavity, with $T_{l(r)}$ being the left (right) mirror transmission and $\tau$ the photon round trip time inside the cavity. In addition, $c^{(\rm in)}_{l}$ and $c^{(\rm in)}_{r}$ are the amplitudes of the left and right driving fields.

In the rotating frame with respect to the frequency  $\omega_r$ of the right driving field, the system Hamiltonian in Eq.~(\ref{eq1}) becomes
\begin{align}
\mathcal{H}=& U^\dag H_{\rm sys} U-i U^\dag \partial_t U\nonumber\\
=&\Delta_c c^\dag c+\Delta_{\rm TLS}\sigma_z+g(c^\dag \sigma_-+c \sigma_+)\nonumber\\
&+i(\mathcal{G} c^{\dag2}-\mathcal{G}^*c^2)+i(\Omega_d c^\dag-\Omega_d^* c),\label{eq5}
\end{align}
where $U=\exp[-i\omega_r(c^\dag c +\sigma_z )t]$ is a unitary transformation operator, $\Delta_c=\omega_c-\omega_r$ is the frequency detuning of the cavity field from the right driving field, and $\Delta_{\rm TLS}=\omega_{\rm TLS}-\omega_r$ is the frequency detuning of the two-level atom from the right driving field. The coupling parameters are given by $\mathcal{G}=Ge^{-i(\omega_p-2\omega_r)t}$ and $\Omega_d=\Omega_l e^{-i(\omega_l-\omega_r)t}+\Omega_r$. Here we choose the right driving field as the reference field, so for convenience we set the phase of the reference field to be zero.

Furthermore, we consider the situation that two external fields are resonant and the frequency of the pumping field on the SOC is twice of that of the right driving field, i.e., $\omega_l=\omega_r$ and $\omega_p=2\omega_r$. The latter condition physically means that a pair of degenerate photons with frequency $\omega_r$ can be obtained when the SOC is illuminated by a field with frequency $\omega_p$. These two conditions directly lead to $\Omega_d=\Omega_l+\Omega_r$ and $\mathcal{G}=G$, respectively. Therefore, the time-dependent Hamiltonian in Eq.~(\ref{eq5}) reduces to
\begin{align}
	H=&\Delta_c c^\dag c+\Delta_{\rm TLS}\sigma_z+g(c^\dag \sigma_-+c \sigma_+)\nonumber\\
	&+i(Gc^{\dag2}-G^*c^2)+i(\Omega_d c^\dag-\Omega_d^* c),\label{eq6}
\end{align}
which is a time-independent Hamiltonian. Note that the above Hamiltonian can be simulated by a superconducting circuit coupled to a nitrogen-vaccancy center in diamond, where the coupling between the cavity and the atom can be amplified exponentially~\cite{wx}.

\section{Steady-state intracavity field}

Using the Heisenberg-Langevin approach, the quantum dynamics of the considered system as described by the Hamiltonian (\ref{eq6}) can be governed by the following quantum Langevin equations:
\begin{align}
\frac{d c(t)}{dt} =& -(\kappa/2+i\Delta_c)c-ig\sigma_-+2Gc^\dagger+\Omega_d+c_{\rm in}(t),\\
\frac{d \sigma_-(t)}{dt} =& -(\gamma/2+i\Delta_{\rm TLS})\sigma_-+2i g c \sigma_z + \sigma_{\rm in}^-(t),\\
\frac{d \sigma_z(t)}{dt} =& -\gamma(\sigma_z+1/2)+ig(c^\dag \sigma_--c\sigma_+)+\sigma_{\rm in}^z(t).
\end{align}
Here, $\kappa=\kappa_l+\kappa_r$ is the total decay rate of the cavity mode, where $\kappa_l$ ($\kappa_r$) is the external decay rate of the left (right) mirror of the cavity, $\gamma$ is the decay rate of the two-level atom. $\kappa$ and $\gamma$ are obtained within Markov approximation, where frequency-dependent decay rates $\kappa[\omega]$ and $\gamma[\omega]$ (or coupling strength $g[\omega]$ between system and bath) are regarded as constants $\kappa$ and $\gamma$. $c_{\rm in}(t)$, $\sigma_{\rm in}^-(t)$ and $\sigma_{\rm in}^z(t)$ are quantum input noises, which depend on the bath operators at initial time. Under Markov approximation, two-time correlation function of these input noises are all written as delta function.   In the steady-state limit, the average values of these input noises and the time-derivatives of the mean values of the system operators vanish, i.e., $\langle c_{\rm in}(t)\rangle=\langle \sigma_{\rm in}^-(t)\rangle=\langle \sigma_{\rm in}^z(t)\rangle=0$ and $d\langle c(t)\rangle/dt=d\langle \sigma_-(t)\rangle/dt=d\langle \sigma_z(t)\rangle/dt=0$. Then, we have the following coupled equations for $\langle c(t)\rangle,~\langle \sigma_-(t)\rangle$ and $\langle \sigma_z(t)\rangle$:
\begin{align}
-(\kappa/2+i\Delta_c)\langle c\rangle-ig\langle \sigma_-\rangle +2G\langle c^\dagger\rangle+\Omega_d =& 0,\label{eq10}\\
 -(\gamma/2+i\Delta_{\rm TLS})\langle \sigma_-\rangle+2i g\langle c\sigma_z\rangle =& 0,\label{eq11}\\
 -\gamma(\langle\sigma_z\rangle+1/2)+ig(\langle c^\dag \sigma_-\rangle-\langle \sigma_+c\rangle) =& 0.\label{eq12}
\end{align}
Using the mean-field approximation, the terms $\langle c\sigma_z\rangle$, $\langle c^\dagger \sigma_-\rangle$ and $\langle \sigma_+c\rangle$ in Eq.~(\ref{eq11}) and Eq.~(\ref{eq12})  can be written respectively as  $\langle c\sigma_z\rangle=\langle c\rangle\langle\sigma_z\rangle$, $\langle c^\dagger \sigma_-\rangle=\langle c^\dagger\rangle \langle\sigma_-\rangle$ and $\langle \sigma_+c\rangle=\langle\sigma_+\rangle \langle c\rangle $. Then the degrees of freedom of the two-level atom can be eliminated by solving Eq.~(\ref{eq10}) and Eq.~(\ref{eq11}), after applying  conjugation. Thus, Eqs.~(\ref{eq10} - \ref{eq12}) can be further reduced to
\begin{align}
-(\kappa_0+i\Delta_0)\langle c\rangle +2G\langle c^\dagger\rangle+\Omega_d =& 0,\label{eq13}\\
-(\kappa_0-i\Delta_0)\langle c^\dag\rangle +2G^*\langle c\rangle+\Omega_d^* =& 0,\label{eq14}
\end{align}
where
\begin{align}
\kappa_0=&\frac{\kappa}{2}+\frac{g^2\gamma/2}{\gamma^2/4+\Delta_{\rm TLS}^2+2g^2n_c},\label{eq15}\\
\Delta_0=&\Delta_c-\frac{g^2\Delta_{\rm TLS}}{\gamma^2/4+\Delta_{\rm TLS}^2+2g^2n_c}.\label{eq16}
\end{align}
Here, $\kappa_0$ and $\Delta_0$ can be interpreted as atom-induced effective cavity linewidth and frequency. Obviously, both depend on the average photon number $n_c=\langle c^\dag c\rangle$ in the cavity. From Eqs.~(\ref{eq13}) and (\ref{eq14}), the steady-state solution of the intracavity field can  easily be obtained as
\begin{align}
\langle c\rangle=\frac{(\kappa_0-i\Delta_0)\Omega_d+2G\Omega_d^*}{\kappa_0^2+\Delta_0^2-4|G|^2}.\label{eq17}
\end{align}
As the average photon number $n_c$ in Eq.~(\ref{eq17}) depends nonlinearly on the system parameters, it may exhibit a bistability as one varies, for example,  the amplitudes of  the driving fields at the mirrors.

\section{CPA criterion}

Below we focus our interest on the dependence of the steady-state output fields on  the driving fields. Using standard input-output theory~\cite{dfw}, the steady-state output fields from the two mirrors of the cavity can be expressed as
\begin{align}
\langle c^{\rm (out)}_l\rangle=&\sqrt{\kappa_l}\langle c\rangle-c^{\rm (in)}_l,\label{eq18}\\
\langle c^{\rm (out)}_r\rangle=&\sqrt{\kappa_r}\langle c\rangle-c^{\rm (in)}_r.\label{eq19}
\end{align}
When CPA occurs,  the input fields are totally absorbed by the coupled atom-cavity system so that $\langle c_{\rm out}^l\rangle=\langle c_{\rm out}^r\rangle=0$. This directly leads to
\begin{align}
c^{\rm (in)}_l/c^{\rm (in)}_r=\sqrt{\kappa_l/\kappa_r},\label{eq20}
\end{align}
It expresses a constraint that the two input fields and the two decay rates of mirrors must satisfy before CPA can be realized. Also, Eq.~(\ref{eq20}) shows that the two input fields must be in phase. For simplicity, $\kappa_l=\kappa_r=\kappa/2$ is assumed in the following. This assumption gives rise to $c^{\rm (in)}_l=c^{\rm (in)}_r=c^{\rm (in)}$ according to Eq.~(\ref{eq20}) and thus also $\Omega_l=\Omega_r=\Omega_d/2$. Without loss of generality, $\Omega_d$ is assumed to be real hereinafter.

Note that the condition in Eq.~(\ref{eq20}) for CPA is necessary but not sufficient. To derive the necessary conditions, we set $\langle c^{\rm (out)}_l\rangle=0$ in Eq.~(\ref{eq18}) [or equivalently $\langle c^{\rm (out)}_r\rangle$=0 in Eq.~(\ref{eq19})]. Then
\begin{equation}
	\sqrt{\kappa/2} \langle c\rangle=c^{\rm (in)},
\end{equation}
i.e.,
\begin{equation}
	\kappa \langle c\rangle=\Omega_d.\label{eq22}
\end{equation}
Eq.~(\ref{eq22}) further gives ${\rm Re}[\langle c\rangle]=\Omega_d/\kappa$ and ${\rm Im}[\langle c\rangle]=0$. Using Eq.~(\ref{eq17}), we obtain 
{\begin{align}
\frac{\kappa_0+2|G|\cos\phi}{\kappa_0^2+\Delta_0^2-4|G|^2}&=\frac{1}{\kappa},\label{1}\\
2|G|\sin\phi&=\Delta_0.\label{2}
\end{align}
Then we replace $\Delta_0$ in Eq.~(\ref{1}) with $2|G|\sin\phi$,  and
\begin{align}
\kappa=\kappa_0-2|G|\cos\phi\label{3}
\end{align}
is given. This equation further gives rise to
\begin{align}
\frac{\kappa/2+2|G|\cos\phi}{\gamma}=\frac{g^2}{2(\gamma^2/4+\Delta_{\rm TLS}^2+2g^2 n_c}.\label{4}
\end{align}
In addition, Eq.~(\ref{2}) can be specifically written as
\begin{align}
\frac{\Delta_c^\prime}{\Delta_{\rm TLS}}=\frac{g^2}{\gamma^2/4+\Delta_{\rm TLS}^2+2g^2 n_c}.\label{5}
\end{align}
Combining Eqs.~(\ref{4}) and (\ref{5}), we obtain the following two equations}
\begin{align}
	\frac{\beta}{\gamma}=&\frac{\Delta_c^\prime}{2\Delta_{\rm TLS}},\label{eq23}\\
	n_c=&\frac{1}{4}(\frac{\gamma}{\beta}-\frac{\gamma^2+4\Delta_{\rm TLS}^2}{2g^2}),\label{eq24}
\end{align}
where $\Delta_c^\prime=\Delta_c-2|G|\sin\phi$ and $G=|G|e^{i\phi}$ with $\phi$ being the relative phase of the pumping field with respect to the reference field, and
\begin{equation}
	\beta=\kappa/2+2|G|\cos\phi.\label{eq25}
\end{equation}
Obviously, the parameter $\beta$ is tunable via  the strength $|G|$ and the relative phase $\phi$ of the pumping field. In addition, conditions in Eqs.~(\ref{eq23}) and (\ref{eq24}) naturally satisfy Eq.~(\ref{eq20}) since they are directly deduced from the condition $\langle c^{\rm (out)}_l\rangle=0 $ (or $\langle c^{\rm (out)}_r\rangle=0 $). Therefore, a necessary condition of CPA is that Eqs.~(\ref{eq23}) and (\ref{eq24}) are simutaneously valid.
By comparing conditions in Eqs.~(\ref{eq23}) and (\ref{eq24}) with conditions obtained in Ref.~\cite{ag1}, the effective CPA Hamiltonian of the system with an effective cavity frequency $\Delta_c^\prime$ can be written as
\begin{align}
H_{\rm eff}=\Delta_c^\prime c^\dag c+\Delta_{\rm TLS}\sigma_z+g(c^\dag \sigma_-+c\sigma_+).
\end{align}

\begin{figure}
	\includegraphics[scale=0.48]{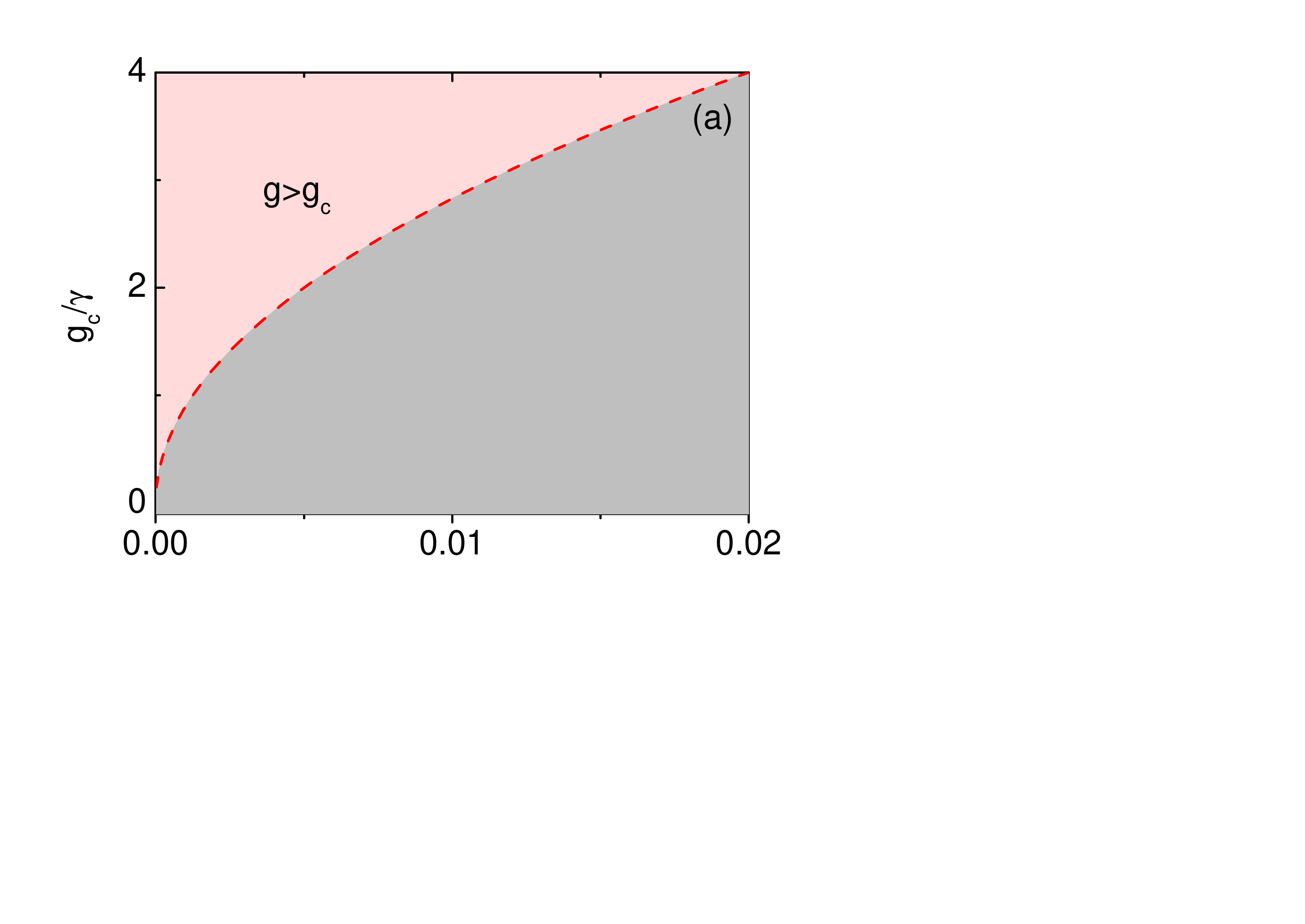}\\
	\includegraphics[scale=0.48]{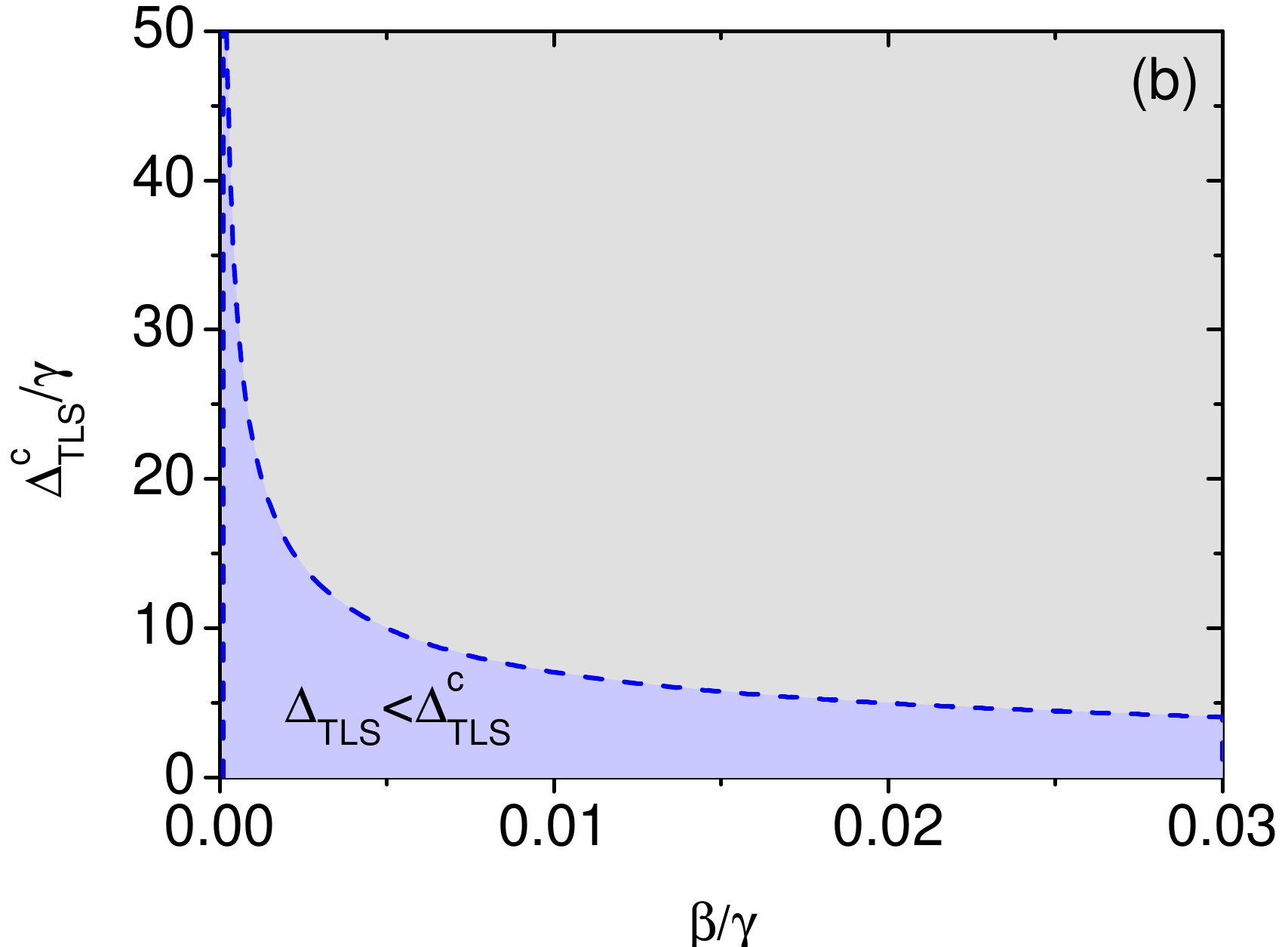}
	\caption{(Color online)The boundary of (a) the coupling strength $g$ between the cavity and the TLS, and (b) the frequency detuning $\Delta_{\rm TLS}$ of the TLS from the driving field vs the tunable parameter $\beta$ in unit of $\gamma$ for prediction of CPA. In (a) $g/\gamma=1$ and (b) $\Delta_{\rm TLS}/\gamma=20$.}\label{fig2}
\end{figure}

Eq.~(\ref{eq24}) also gives a constraint on the coupling strength $g$ between the two-level system and the cavity for any given value of the detuning $\Delta_{\rm TLS}$ [see Fig.~\ref{fig2}(a)]. Specifically, the mean intracavity photon number is positive, i.e., $n_c>0$. Hence
one requires that
\begin{align}
g >g_c(\beta,\Delta_{\rm TLS},\gamma)\equiv\sqrt{\frac{1}{2}\beta(\gamma+4\Delta_{\rm TLS}^2/\gamma)}\label{eq26}
\end{align}
where $g_c(\beta,\Delta_{\rm TLS},\gamma)$ is a critical value of $g$ for the occurrence of CPA. Therefore, the condition in Eq.~(\ref{eq26}) can be satisfied by fine-tuning the parameters $\beta$ and $\Delta_{\rm TLS}$ for any given $\gamma$. In particular, for $\Delta_{\rm TLS}=0$, i.e. when the two-level atom is resonant with the input field, Eq.~(\ref{eq26}) becomes
\begin{align}
g>g_c(\beta,\gamma)=\sqrt{\beta\gamma/2}.\label{eq27}
\end{align}
This gives a minimum value of the coupling strength $g$ for CPA to occur. Obviously, the requirement in Eq.~(\ref{eq27}) is more flexible than in previous studies~\cite{ag1,ag2,ag3} because of the introduction of the tunable parameter $\beta$. Without the second-order nonlinearity ($G=0$),  Eq.~(\ref{eq25}) reduces to $\beta=\kappa/2$. According to Eq.~(\ref{eq27}), $g>g_c(\kappa,\gamma)=\sqrt{\kappa\gamma/2}$ is required to generate CPA, which satisfies $g^2/\kappa\gamma>1$ and corresponds to the strong coupling regime. Such a strong coupling can be realized by replacing a single two-level atom with an atomic ensemble, which has been studied in previous works \cite{ag1,ag2,ag3}. In sharp contrast, with the second-order nonlinearity ($G\neq0$), the parameter $\beta$ can be tuned to an extremely small value via appropriate choices of $G$ and $\phi$. For example, the relative phase $\phi$ (or $\cos\phi$) can be chosen so that $\beta=0.01\gamma$ for any given $G$. This value is much smaller than the decay rate $\kappa$ of the cavity for pratical atom/spin-cavity systems since $\gamma<\kappa$. Then the condition in Eq.~(\ref{eq27}) becomes $g^2/\kappa \gamma>0.01$. CPA can thus occur over a wide parameter range satisfying $g^2/\kappa\gamma<1$, corresponding to the weak coupling regime. This shows that CPA can indeed occur in the weak coupling regime for our setup. At present, realizing a strong coupling between  a single two-level system (e.g., a nitrogen vacancy center in diamond) and a cavity  or a superconducting circuit~\cite{wx} is still a challenge~\cite{p.lodahl}. Therefore, exploring optical phenomena in weakly coupled quantum system is of great significance.

Eq.~(\ref{eq24}) does not only limit the coupling strength $g$ for the occurrence of CPA, but also gives a constraint on the detuning $\Delta_{\rm TLS}$. For a given $g$, CPA can only be observed when $|\Delta_{\rm TLS}|<\sqrt{\frac{1}{2}(\frac{g^2\gamma}{\beta}-\frac{\gamma^2}{2})}\equiv|\Delta_{\rm TLS}^c|$ [see the light blue region in Fig.~\ref{fig2}(b)]. The dashed blue curve represents the critical detuning $\Delta_{\rm TLS}^c$ against the parameter $\beta$. From Fig.~\ref{fig2}(b), we see that $\Delta_{\rm TLS}$ can vary in a broader range than in Ref.~\cite{ag2} for a fixed $\kappa$. This results from the introduction of the {\it controllable} parameter $\beta$. As mentioned above, $\beta$ is allowed to be very small, so a large $\Delta_{\rm TLS}$ is needed, leading to a small mean photon number $n_c$ according to Eq.~(\ref{eq24}). Therefore, weak input fields are sufficient to achieve CPA with our setup. This greatly simplifies experimental implementations.

\begin{figure}
	\centering
\includegraphics[scale=0.48]{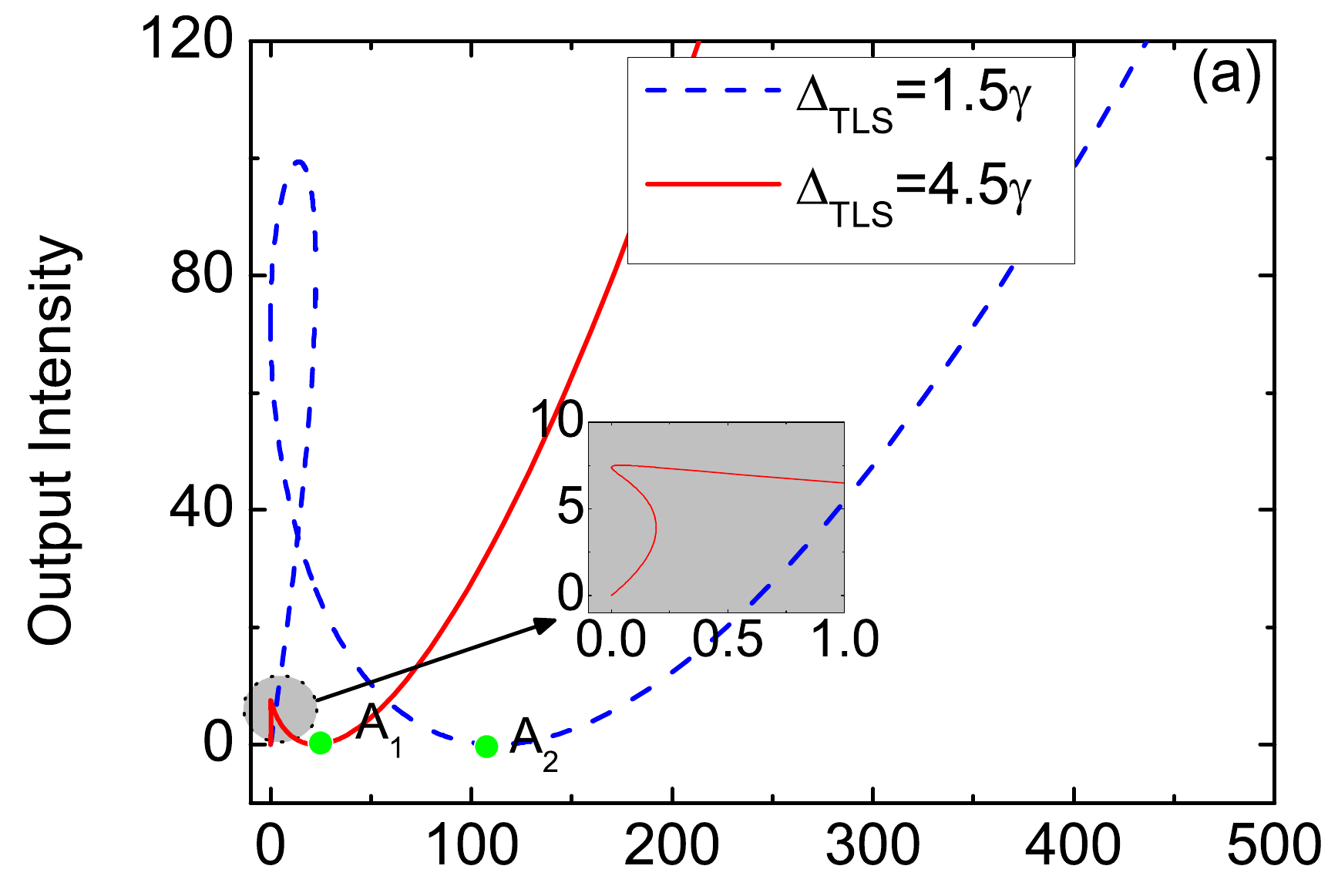}
\includegraphics[scale=0.48]{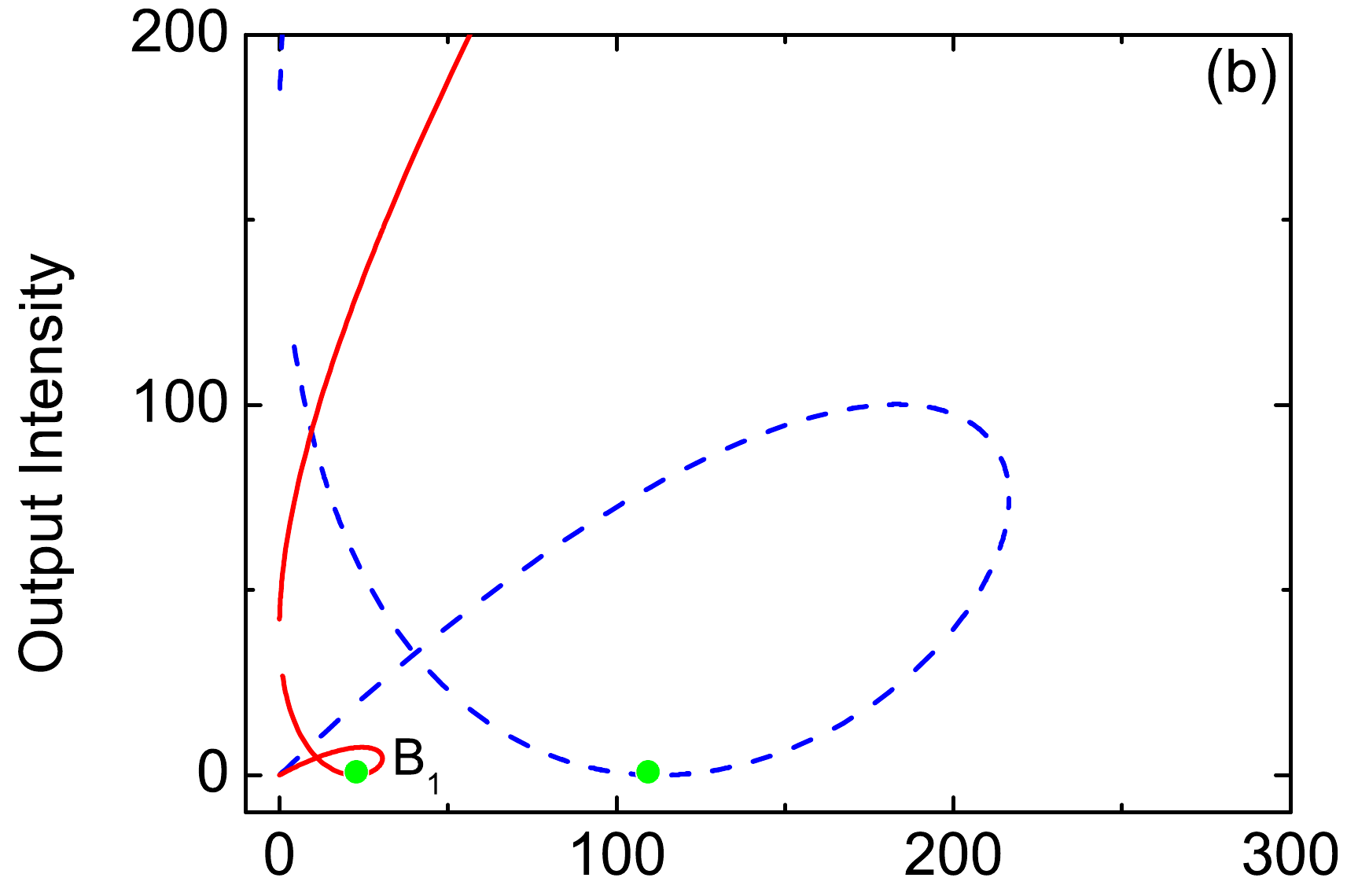}
\includegraphics[scale=0.48]{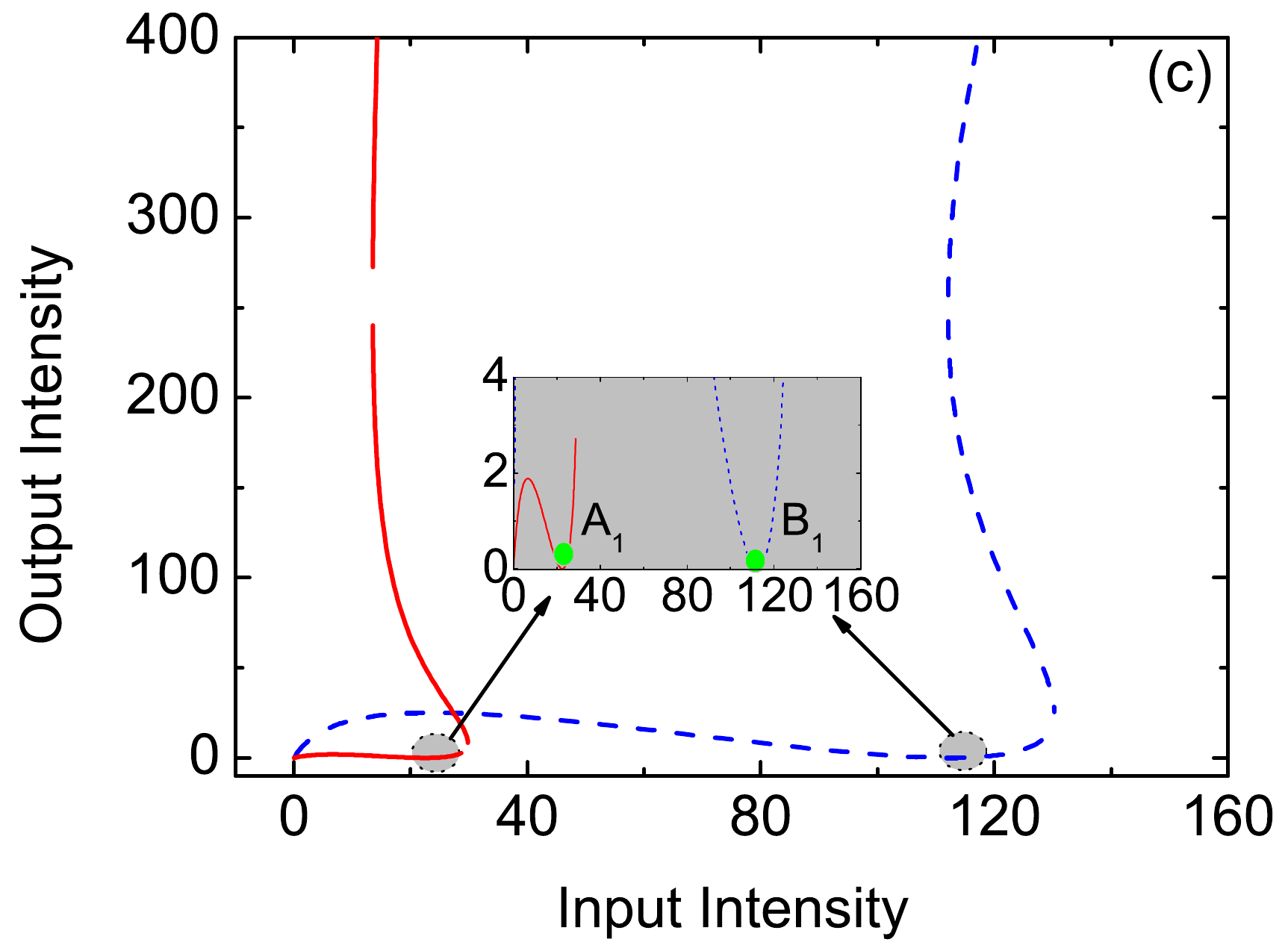}
\caption{(Color online) The output intensity as a function of the input intensity with $\Delta_{\rm TLS}=4.5\gamma$ and $1.5\gamma$. The relative phase is (a) $\phi=2/3\pi$, (b) $\phi=4/3\pi$, and (c) $\phi=\pi$.}\label{fig3}
\end{figure}

\section{numerical results}

We now numerically study CPA based on our setup in the weak coupling regime.  We put $g=\gamma,~\kappa=20\gamma$, and $\beta=0.02\gamma$, so that $g^2/\kappa\gamma<1$, $g^2/\beta\gamma>1$ and $\Delta_{\rm TLS}^c\approx4.975\gamma$. This indicates that CPA can only be observed for $|\Delta_{\rm TLS}|\in[0,4.975)$. As CPA can only occur when both Eqs.~(\ref{eq23}) and (\ref{eq24}) are simultaneously satisfied, the parameter $\Delta_c$ can be solved in particular using Eq.~(\ref{eq23}). Adopting these parameters, we numerically solve Eqs.~(\ref{eq17}) to obatin the average photon number $n_c$, which is then substituted into Eqs.~(\ref{eq18}) and (\ref{eq19}) to obtain the output intensity $|c^{\rm (out)}|^2$. Fig.~\ref{fig3} plots the output intensity as a function of the input intensity $|c^{\rm (in)}|^2$ for frequency detuning $\Delta_{\rm TLS}=4.5 \gamma$ and $1.5 \gamma$ under various conditions.

Fig.~\ref{fig3}(a) shows results for $|G|=9.98\gamma$ and $\phi=2/3\pi$. We observe that the output intensity exhibits a bistability with respect to the input intensity. For Eq. (\ref{eq17}), a certain value of the input field is given, two simultaneously stable fixed-point solutions are obtained, which gives the bistability of the output field according to the input-output theory. Also, by increasing and decreasing the input field, the hysteresis curve is obtained. The bistablility is closely related to bifurcation phenomena \cite{pa,lintian}. The critical points of bistability can be regarded as two bifurcation points \cite{lintian}

By varying the parameter $\Delta_{\rm TLS}$, the bistable pattern can be changed from a conventional [inset in Fig.~\ref{fig3}(a)] to an unconventional [ dashed blue curve in Fig.~\ref{fig3}(a)] shape. When the average intracavity photon number satisfies Eq.~(\ref{eq24}) corresponding to the green dots $A_1$ and $A_2$, CPA is predicted. They are both located at the upper branch of bistable pattern. Therefore, the CPA conditions are outside the bistable region.

By vary the relative phase $\phi$ from $2/3\pi$ to $4/3\pi$ as shown in Fig.~\ref{fig3}(b), the bistable pattern of the output intensity with respect to the input intensity becomes robust against variations of the TLS frequency detuning $\Delta_{\rm TLS}$ and takes the unconventional shape. Interestly, CPA points (see the green dot $B_1$ and $B_2$) appear inside the bistable region and are located at the unstable branches. The different locations of CPA points in Figs.~\ref{fig3}(a) and \ref{fig3}(b) are caused by the different values of $\Delta_c'$ from Eq.~(\ref{eq23}) resulting from  the different  $\phi$ considered. However, CPA in Fig.~\ref{fig3}(b) cannot be observed experimentally due to the unstable nature.

We further study the case of $\phi=\pi$ and results are shown in Fig.~\ref{fig3}(c). To ensure $\beta=0.02\gamma$, we have considered an decreased amplitude $|G|=4.99\gamma$ of the pumping field on the SOC. The output intensity also exhibits a bistable behavior with respect to the input intensity with convectional bistable patterns for both values of $\Delta_{\rm TLS}$ studied. In addition, the CPA points are located inside the bistable regime in the stable branches [see points $A_1$ and $B_1$ in inset of Fig.~\ref{fig3}(c) ].

Fig.~\ref{fig3} shows that the CPA points are blue-shifted upon decreasing the frequency detuning of the TLS. Also, their locations are unaffected by the parameters $|G|$ and $\phi$ for fixed $\Delta_{\rm TLS}$ [see red and dashed blue curves in Fig.~\ref{fig3}]. Hence, the location of CPA point is robust against the parameters of the SOC. These results also follow directly from Eq.~(\ref{eq24}).

We emphasize differences between our system for realizing CPA from the previously  studied ones~\cite{ag1,ag2,ag3,sl2011}. First, CPA can occur in our case in the weak coupling regime and the system exhibits bistable behaviors. Second, we have shown that the bistable pattern can be changed from the conventional to unconventional shape and CPA point can be tuned to appear at the upper or lower stable branch or even at the unstable branch. Third, the location of CPA point is robust against the parameters of the SOC when the effective decay rate $\beta$ is fixed and it can be either inside or outside the bistable region.

\section{Conclusion and Discussion}

\begin{figure}
	\centering
	\includegraphics[scale=1.2]{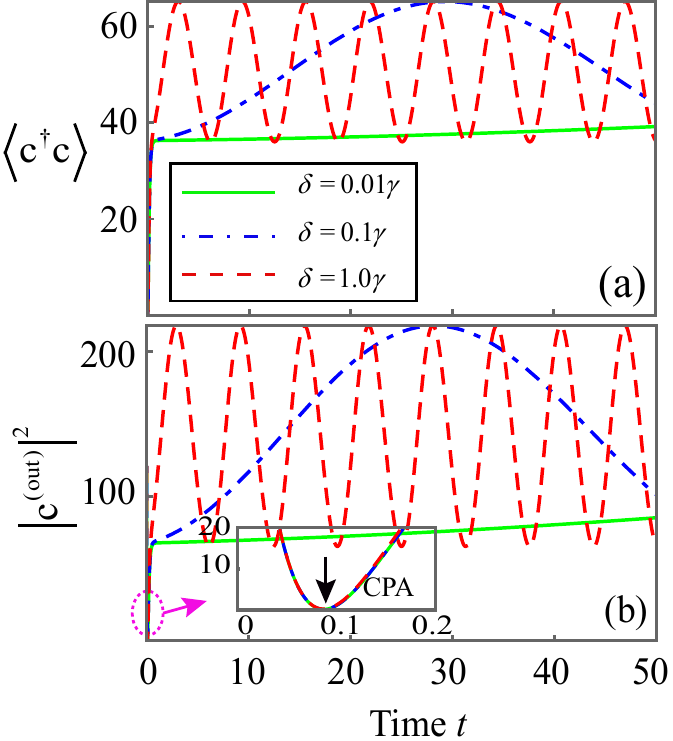}
	\caption{(Color online) The mean photon number and the output intensity versus the evolution time with different frequency detunings $\delta=\omega_p-2\omega_r=0.01\gamma,0.1\gamma,1.0\gamma$ in both (a) and (b).}\label{fig4}
\end{figure}

Note that the condition $\omega_p=2\omega_r$ is used to obtain the time-independent Hamiltonian in Eq.~(\ref{eq6}). This condition seems a little strict. Fortunately, it has been achieved in various experiments using degenerate parametric amplification \cite{jingzhang,xguo}.  For more realistic, we consider the effect induced by the fluctuation $\delta=\omega_p-2\omega_r$ between $\omega_p$ and $2\omega_r$ on the output intensity and the mean photon number $\langle c^\dag c\rangle=n_c$. In Fig.~\ref{fig4}, we show that $n_c$ rapidly increases to a certain value and then oscillates with evolution time [see Fig.~\ref{fig4}(a)], while the output intensity first decreases to a CPA point and then it grows rapidly and oscillates with time evolution [see Fig.~\ref{fig4}(b)]. It is not difficult to find that CPA point is robust against the fluctuation $\delta$ with time evolution for out considered situation.

In summary, we have given a detailed study on CPA in a two-level atom weakly coupled to a cavity embedded with  a SOC. Under CPA conditions, the system behaves as a two-level system coupled to a cavity with a tunable effective bandwidth. The output field intensity  exhibits a bistability. By tuning system parameters such as the frequency detuning of two-level atom from the input driving field, the coefficient of the nonlinearity crystal and its relative phase, the bistable pattern can be switched from conventional to unconventional sharps or vice versa. Due to the effect of the SOC on the effective frequency $\Delta_c^\prime$, the operation point of CPA can be switched between two stable branches. However, the location of CPA point is robust against the parameters of the SOC when the effective decay rate $\beta$ is fixed. Our study provides a novel way to realize future optical devices utilizing CPA in the weak  coupling regime.

\section*{Acknowledgments}

This work is supported by the National Key Research and Development Program of China (Grant No. 2016YFA0301200), the National Natural Science Foundation of China (Grants No. 11804074, No. 11934010 and No. U1801661), Hong Kong General Research Fund (Grant No. 153017/17P), Talent Development Funding of Hefei University (Grant No. 18-19RC60), Candidates of Hefei University Academic Leader(Grant No.2016dtr02), Major Project of Hefei Preschool Education College (Grant No.hyzzd2018003).




\begin{thebibliography}{99}
	\bibitem{siegmann} A. E. Siegmann, {\it Lasers} (University Science Books, Mill Valley, CA, 1986).	
	
	\bibitem{ydc1}Y. D. Chong, L. Ge, H. Cao, and A. D. Stone, Coherent perfect absorbers: Time-reversed lasers, Phys. Rev. Lett. {\bf 105}, 053901 (2010).
	
	\bibitem{ww} W. Wan, Y. Chong, L. Ge, H. Noh, A. D. Stone, and H. Cao, Time-reversed lasing and interferometric control of absorption, Science {\bf 331}, 889 (2011).
	
	\bibitem{hn}H. Noh, Y. Chong, A. D. Stone, and H. Cao, Perfect coupling of light to surface plasmons by coherent absorption, Phys. Rev. Lett. {\bf 108}, 186805 (2012).
	
	\bibitem{ads}A. D. Stone, Gobbling up light with an antilaser, Phys. Today {\bf 64}, 68 (2011).
	
	\bibitem{mk}M. Kang and Y. D. Chong, Coherent optical control of polarization with a critical metasurface, Phys. Rev. A {\bf 92}, 043826 (2015).
	
	\bibitem{mc}M. Crescimanno, C. Zhou, J. H. Andrews, and M. A. Baker, Structure and symmetry in coherent perfect polarization rotation, Phys. Rev. A {\bf 91}, 013845 (2015).
	
	\bibitem{sdg} S. Dutta-Gupta, R. Deshmukh, A. V. Gopal, O. J. F. Martin, and S. D. Gupta, Coherent perfect absorption mediated anomalous reflection and refraction, Opt. Lett. {\bf 37}, 4452 (2012).
	
	\bibitem{gsd} G. S. Dutta, Strong-interaction mediated critical coupling at two distinct frequencies, Opt. Lett. {\bf 32}, 1483 (2007).
	
	\bibitem{jwy} J. W. Yoon, G. M. Koh, S. H. Song, and R. Magnusson, Measurement and modeling of a complete optical absorption and scattering by coherent surface plasmon-polariton excitation using a silver thin-film grating, Phys. Rev. Lett. {\bf 109}, 257402 (2012).
	
	\bibitem{sl}S. Longhi, PT-symmetric laser absorber, Phys. Rev. A {\bf 82}, 031801(R) (2010).
	
	\bibitem{sl2}S. Longhi, Time-reversed optical parametric oscillation, Phys. Rev. Lett. {\bf 107}, 033901 (2011).
	
	\bibitem{ydc2} Y. D. Chong, L. Ge, and A. D. Stone, PT-symmetry breaking and laser-absorber modes in optical scattering systems, Phys. Rev. Lett. {\bf 106}, 093902 (2011).
	
	\bibitem{ydc3}Y. D. Chong and A. D. Stone, Hidden black: Coherent enhancement of absorption in strongly scattering media, Phys. Rev. Lett. {\bf 107}, 163901(2011).
	
	\bibitem{sh} S. Huang and G. S. Agarwal, Coherent perfect absorption of path entangled single photons, Opt. Express {\bf 22}, 20936 (2014).
	
	\bibitem{jt} J. T. Shen and S. Fan, Quantum critical coupling conditions for zero single-photon transmission through a coupled atom-resonator-waveguide system, Phys. Rev. A, {\bf 82}, 021802 (2010).
	
	\bibitem{xby} X. B. Yan, C. L. Cui, K. H. Gu, X. D. Tian, C. B. Fu, and J. H. Wu, Coherent perfect absorption, transmission, and synthesis in a double-cavity optomechanical system, Opt. Express, {\bf 22}, 4886 (2014).
	
	\bibitem{yml2018}Y. M. Liu, X. D. Tian, J. Wang, C. H. Fan, F. Gao, and Q. Q. Bao, All-optical transistor based on Rydberg atom-assisted optomechanical system, Optics Express, {\bf26}, 12330 (2018).
	
	\bibitem{ag1}G. S. Agarwal and Y. Zhu, Photon trapping in cavity quantum electrodynamics, Phys. Rev. A {\bf 92}, 023824 (2015).
	
	\bibitem{ag2}G. S. Agarwal, K. Di, L. Wang, and Y. Zhu, Perfect photon absorption in the nonlinear regime of cavity quantum electrodynamics, Phys. Rev. A {\bf 93}, 063805 (2016).
	
	\bibitem{ag3}L. Wang, K. Di, Y. Zhu, and G. S. Agarwal, Interference control of perfect photon absorption in cavity quantum electrodynamics, Phys. Rev. A {\bf 95}, 013841 (2017).
	
	\bibitem{tr} T. Roger, S. Vezzoli, E. Bolduc, J. Valente, J. J. F. Heitz, J. Jeffers, C. Soci, J. Leach, C. Couteau, N. I. Zheludev, and D. Faccio, Coherent perfect absorption in deeply subwavelength films in the single-photon regime, Nat. Commun. {\bf 6}, 7031 (2015).
	
	
	\bibitem{yz}Y. Zhang, A. Sohail and C. Yu, Europhysics Letters {\bf 115}, 64002 (2016).
	
	\bibitem{ag4}G. S. Agarwal and S. Huang,  New J. Phys. {\bf 16}, 033023 (2014).
	
	\bibitem{dkz}D. Zhang, X. Q. Luo, Y. P. Wang, T. F. Li, and J. Q. You, Observation of the exceptional point in cavity magnon-polaritons, Nat. Commun. {\bf 8}, 1368 (2017).
	
	\bibitem{sl2011}S. Longhi, Coherent perfect absorption in a homogeneously broadened two-level medium, Phys. Rev. A {\bf 83}, 055804 (2011).
	
	\bibitem{hchoi}H. Choi, M. Heuck, and D. Englund, Self-Similar Nanocavity Design with Ultrasmall Mode Volume for Single-Photon Nonlinearities, Phys. Rev. Lett. {\bf 118}, 223605 (2017).
	
	
	\bibitem{vr1}V. R. Almeida and M. Lipson, Optical bistability on a silicon chip, Opt. Lett. {\bf29}, 2387 (2004).
	
	\bibitem{vr2}V. R. Almeida, C. A. Barrios, R. R. Panepucci, M. Lipson, M. A. Foster, D. G. Ouzounov, and A. L. Gaeta, All-optical switching on a silicon chip, Opt. Lett. {\bf29}, 2867 (2004).
	
   \bibitem{liy}Y. Li, A.Sinitskii, and J. M. Tour, Electronic two-terminal bistable graphitic memories, Nat. Mater {\bf7}, 966 (2008).
	
	\bibitem{cr}R. Cerna, Y. L$\acute{e}$ger, T. K. Paraïso, M. Wouters, F. MorierGenoud, M. T. Portella-Oberli, and B. Deveaud, Ultrafast tristable spin memory of a coherent polariton gas, Nat. Commun. {\bf4}, 2008 (2013).
		
	\bibitem{lr}R. Labouvie, B. Santra, S. Heun, and H. Ott, Bistability in a driven-dissipative superfluid, Phys. Rev. Lett. {\bf116}, 235302(2016). 
	
	\bibitem{lintian} L. Tian, Cavity-assisted dynamical quantum phase transition at bifurcation points, Phys. Rev. A {\bf 93}, 043850 (2016).
	
	\bibitem{vmy}M. Y. Vilensky, Y. Prior, and I. Sh. Averbukh, Cooling in a bistable optical cavity,	Phys. Rev. Lett. {\bf99}, 103002 (2007).
	
	\bibitem{gq}G. Q. Ge, X. Luo, Y. Wu, and Z. Li, Atomic coherence and bistable lasers without inversion, Phys. Rev. A {\bf54}, 1604 (1996).
	
	
	\bibitem{tch}T. C. H. Liew, A. V. Kavokin, and I. A. Shelykh, Optical circuits based on polariton neurons in semiconductor microcavities, Phys. Rev. Lett. {\bf 101}, 016402 (2008). 
	
	\bibitem{teo}T. Espinosa-Ortega and T. C. H. Liew, Complete architecture of integrated photonic circuits based on and and not logic gates of exciton polaritons in semiconductor microcavities, Phys. Rev. B {\bf 87}, 195305 (2013).
	
	\bibitem{ds}D. Sanvitto, S. Pigeon, A. Amo, D. Ballarini, M. De Giorgi, I. Carusotto, R. Hivet, F. Pisanello, V. G. Sala, P. S. S. Guimaraes, R. Houdré, E. Giacobino, C. Ciuti, A. Bramati, and G. Gigli, All-optical control of the quantum flow of a polariton condensate, Nat. Photonics {\bf 5}, 610 (2011).
	
	\bibitem{bd}D. Ballarini, M. De Giorgi, E. Cancellieri, R. Houdr$\acute{e}$, E. Giacobino, R. Cingolani, A. Bramati, G. Gigli, and D. Sanvitto, All-optical polariton transistor, Nat. Commun. {\bf4}, 1778 (2013).
	
	\bibitem{ls}S. Longhi, Non-reciprocal transmission in photonic lattices based on unidirectional coherent perfect absorption, Opt. Lett. {\bf40}, 1278 (2015).

	\bibitem{zu0} M. O. Scully, M. S. Zubairy, {\it Quantum optics}  (Cambridge University, United Kingdom, 1997).
	\bibitem{zu1}  J. Anwar, M. S. Zubairy, Effect of squeezing on the degenerate parametric oscillator, Phys. Rev. A  {\bf 45}, 1804  (1992).
	\bibitem{zu2} K. Wodkiewicz and M. S. Zubairy, Effect of laser fluctuations on squeezed states in a degenerate parametric amplifier, Phys. Rev. A {\bf 27}, 2003 (1983).
		
	\bibitem{wx}W. Xiong, Y. Qiu, L. A. Wu, and J. Q. You, Amplification of the coupling strength in a hybrid quantum system, New J. Phys. {\bf20}, 043037 (2018).
	
	\bibitem{wq} W. Qin, A. Miranowicz, P. B. Li, X. Y. L\"{u}, J. Q. You, and F. Nori, Exponentially-Enhanced Light-Matter Interaction, Cooperativities, and Steady-State Entanglement Using Parametric Amplification, Phys. Rev. Lett. {\bf120}, 093601 (2018).
	
    \bibitem{xylu}X. Y. L\"{u}, Y. Wu, J. R. Johansson, H. Jing, J. Zhang, and F. Nori, Squeezed optomechanics with phase-matched
	amplification and dissipation, Phys. Rev. Lett {\bf 114},
	093602 (2015).
	
	\bibitem{dfw}D. F. Walls and G. J. Milburn, {\it Quantum optics} (Springer-Verlag, Berlin, 1994).
	
	\bibitem{p.lodahl}P. Lodahl, S. Mahmoodian, and S\o ren Stobbe, Interfacing single photons and single quantum dots with photonic nanostructures, Rev. Mod. Phys. {\bf87}, 347 (2015).
	
	\bibitem{pa} A. Patra, B. L. Altshuler, and E. A. Yuzbashyan, Phys. Rev. A {\bf 99}, 033802 (2019).
	
	
	\bibitem{jingzhang}H. Ma, C. Ye, D. Wei, and J. Zhang, Coherence phenomena in the phase-sensitive optical parametric amplification inside a cavity, Phys. Rev. Lett{\bf 95}, 233601 (2005).

	\bibitem{xguo}X. Guo, C. Zou, C. Schuck, H. Jung, R. Cheng and H. Tang, Parametric down-conversion photon-pair source on a nanophotonic chip, Light: Science \& Applications {\bf 6}, e16249 (2017).
\end{thebibliography}
\end{document}